\documentclass[preprint,showpacs,preprintnumbers,amsmath,amssymb,pra]{revtex4}
\usepackage{graphicx} 
\usepackage{amsmath}
\usepackage{dcolumn}
\usepackage{bm}
\numberwithin{equation}{section} 
\newcommand{\bra}[1]{\langle#1|}

\newcommand{\ket}[1]{|#1\rangle} 
\newcommand{\braket}[2]{\langle#1|#2\rangle} 

\newcommand{\me}[3]{{#2}_{#1,#3}}

\newcommand{\EQ}[1]{Eq. (\ref{#1})}
\usecounter{figure}

\begin{document}

\title{The augmented message-matrix approach to deterministic dense coding theory}

\author{E.~Gerjuoy}
\affiliation{Department of Physics, University of Pittsburgh, Pittsburgh, PA 15260.}
\email{gerjuoy@pitt.edu}
\author{H.~T.~Williams}
\affiliation{Department of Physics and Engineering, Washington and Lee University, Lexington, VA 24450}
\email{williamsh@wlu.edu}
\author{P.~S.~Bourdon}
\affiliation{Department of Mathematics, Washington and Lee University, Lexington, VA 24450} 
\email{pbourdon@wlu.edu}
\date{\today}

\begin{abstract}
  A method is presented for producing analytical results applicable to the standard two-party deterministic dense coding protocol, wherein communication of $K$ perfectly distinguishable messages is attainable with the aid of $K$ selected local unitary operations on one qudit from a pair of entangled qudits of equal dimension $d$ in a pure state $\ket{\psi}$ with largest Schmidt coefficient $\sqrt{\lambda_0}$.  The method utilizes the properties of a $d^2 \! \times \! d^2$ unitary matrix whose initial columns represent message states of the system used for communication, augmented by sufficiently many additional orthonormal column vectors so that the resulting matrix is unitary.  Using the unitarity properties of this augmented message-matrix, we produce simple proofs of previously established results including (i) the bound $\lambda_0 \leq d/K$, and (ii) the impossibility of finding a $\ket{\psi}$ that can enable transmission of $K=d^2-1$ messages but not $d^2$.  Additional results obtained using the method include proofs that when $K=d+1$ the $\lambda_0 \leq d/K$ bound (i) always reduces to at least $\lambda_0 \leq (1/2)[1+\sqrt{(d-2)/(d+2)}]$, and (ii) reduces to $\lambda_0 \leq (d-1)/d$ in the special case that the identity and shift operators are two of the selected local unitaries.
\end{abstract}

\pacs{03.65.Ud, 03.67.Hk}
\maketitle

\section{\label{iandf}Introduction and Formalism\protect\\}
The deterministic dense coding protocol, first described by Bennett and Wiesner \cite{BW} in 1992, has been the subject of numerous investigations (\cite{M}--\cite{B} and references therein).  Here, therefore, we describe the protocol and its associated formalism only briefly.

Alice and Bob, who are located far apart, each controls one qudit from an entangled pair.  Orthonormal basis sets for Alice's and Bob's qudits, in their respective Hilbert spaces $H_A$ and $H_B$, are denoted respectively by $\ket{i}_A$ and $\ket{j}_B$, $i,j=0,1,\ldots,(d-1)$.  Initially the pair of qudits is in a normalized entangled pure state $\ket{\psi}$, with Schmidt representation
\begin{equation}\label{EG1}
\ket{\psi} = \sum_{j=0}^{d-1}\sqrt{\lambda_j}\, \ket{j}_A\ket{j}_B \equiv \sum_{j=0}^{d-1} \sqrt{\lambda_j}\, \ket{jj} .
\end{equation}
In Eq.~(\ref{EG1}) the Schmidt coefficients $\sqrt{\lambda_j}$ are non-negative real numbers satisfying $\sum_{j=0}^{d-1} \lambda_j = 1$; we make the conventional assumption, without loss of generality, that $\lambda_0 \geq \lambda_1 \geq \ldots \geq \lambda_{d-1} \geq 0$.  The right side of Eq.~(\ref{EG1}) makes use of the convenient notation, which we shall employ henceforth, that $\ket{ij}$ denotes the product basis state $\ket{i}_A \ket{j}_B$; collectively these states form a complete orthonormal basis set in the $d^2$-dimensional Hilbert space $H = H_A \otimes H_B$, wherein lie $\ket{\psi}$ and all other state functions describing the state of the qudit pair.  Alice performs a local unitary operation $U_A$ on her qudit and then sends the qudit to Bob via a noise-free quantum channel.  Any such $U_A$ converts $\ket{\psi}$ to the normalized state function
\begin{equation} \label{EG2}
  \ket{\Psi} = (U_A \otimes I_B) \ket{\psi} = \sum_{i,j=0}^{d-1} \sqrt{\lambda_j}U_{ij} \ket{ij},
\end{equation}
where $U_{ij}$ denotes the matrix element $\bra{i} U_A \ket{j}$ of the operator $U_A = \sum_{i,j=0}^{d-1} U_{ij} \ket{i}_A \bra{j}_A$.  Let $\{U^{(a)}\}_{a=0}^{K-1}$ be a set of $K$ local unitaries having the special property that the $K$ corresponding $\ket{\Psi^{(a)}}$'s are mutually orthogonal, with $K$ here and hereinafter the largest possible number of such unitaries for a given $\ket{\psi}$.  As Mozes \textit{et al.}~\cite{M} have observed, the condition that the $U^{(a)}$ constitute such a set is expressed by the requirement that, for every $a,b$ pair in the set,
\begin{equation} \label{EG3}
  \braket{\Psi^{(a)}}{\Psi^{(b)}} \equiv \sum_{i,j=0}^{d-1} \lambda_j (U_{ij}^{(a)})^* U_{ij}^{(b)} = \text{tr} ( \Lambda (U^{(a)})^{\dagger} U^{(b)}) = \delta_{ab},
\end{equation}
where $\braket{\Psi^{(a)}}{\Psi^{(b)}}$ denotes the Hilbert space $H$ scalar product of $\ket{\Psi^{(a)}}$ and $\ket{\Psi^{(b)}}$, and $\Lambda$ is a diagonal $d \times d$ matrix whose diagonal elements are the squares of the Schmidt coefficients defined in Eq.~(\ref{EG1}), i.e., $\Lambda_{ij} = \lambda_i \delta_{ij}$.  A set $\{U^{(a)}\}$ satisfying Eq.~(\ref{EG3}) will be termed "$\Lambda$ orthogonal."

If Bob knows Alice has operated on $\ket{\psi}$ with one of the $K$ unitaries in some given $\Lambda$-orthogonal  set $\{U^{(a)}\}$, then Bob---after receiving Alice's qudit---can correctly determine which particular $U^{(a)}$ Alice actually employed before sending her qudit.  Thus this protocol enables Alice to send Bob one of $K$ previously agreed-upon possible messages.  Deterministic dense coding theory seeks to answer the question:  Given specified values for the Schmidt coefficients, what is the corresponding value of $K$?  It has been demonstrated \cite{BW} that when every $\lambda_i = 1/d$ in Eq.~(\ref{EG1}), i.e. when $\ket{\psi}$ is maximally entangled, then $K=d^2$.  For non-maximally entangled $\ket{\psi}$, however, tantalizing open questions remain about the dependence of $K$ on the $\lambda_i$, despite significant scrutiny given to dense coding theory.  In particular, Mozes \textit{et al.} \cite{M} have numerically explored this dependence in great detail for the three-dimensional case ($d=3$), and in lesser but still illuminating detail for $d=4$ to $7$.  In so doing they produced several interesting conjectures which will be examined among the issues treated herein.

\subsection{\label{tpc}This Paper's Contributions}

Given any set of $K$ unitaries $\{U^{(a)}\}_{a=0}^{K-1}$ constituting a $\Lambda$-orthogonal set, Eq.~(\ref{EG3}) shows the corresponding $\{\Psi^{(a)}\}_{a=0}^{K-1}$ can be thought of as a set of $K$ orthonormal basis vectors for the $K$-dimensional subspace $S_K$ of the $d^2$-dimensional Hilbert space $H$.  Any set of $d^2-K$ orthonormal basis vectors $\{\Phi^{(b)}\}_{b=K}^{d^2-1}$, contained in the subspace $S_{d^2-K}$ ortho-complementary to $S_K$, must be orthogonal to the set $\{\Psi^{(a)}\}_{a=0}^{K-1}$.  Observe that $\{\Psi^{(a)}\}_{a=0}^{K-1}$ taken together with $\{\Phi^{(b)}\}_{b=K}^{d^2-1}$ is an orthonormal basis for $H$.  Of course, both $\ket{\Psi^{(a)}}$ and $\ket{\Phi^{(b)}}$ can be expressed in terms of their components along elements of the basis $B = \{\ket{ij}: 0 \leq i, j \leq d-1\}$:
\begin{equation} \label{PB03}
  \ket{\Psi^{(a)}} = \sum_{i,j=0}^{d-1} \sqrt{\lambda_j} U_{ij}^{(a)} \ket{ij} \text{ and } \ket{\Phi^{(b)}} = \sum_{i,j=0}^{d-1} \phi_{ij}^{(b)} \ket{ij} .
\end{equation}
We order the basis $B$ of $H$ in a nonstandard way as follows
\begin{equation} \label{PB04}
   B = (\ket{00},\ket{10}, \ldots,\ket{(d-1)0},\ket{01},\ket{11},\ldots,\ket{(d-1)1},\ldots,\ket{0(d-1)},\ket{1(d-1)},\ldots,\ket{(d-1)(d-1)}).
\end{equation}

Central to our approach is the $d^2 \times d^2$ matrix $M$, whose entries are the components of the vectors $\ket{\Psi^{(a)}}$ and $\ket{\Phi^{(b)}}$ relative to $B$.  For $0 \leq a \leq K-1$, column $a$ of $M$ comprises the components of $\ket{\Psi^{(a)}}$ relative to $B$.  For $K \leq b \leq d^2-1$, column $b$ of $M$ comprises the components of $\ket{\Phi^{(b)}}$  relative to $B$.  We refer to $M$ as an \textit{augmented message-matrix} since it is composed of column vectors representing the two-qudit states $\ket{\Psi^{(a)}}$ that Alice can prepare as messages for Bob, augmented with enough additional vectors $\ket{\Phi^{(b)}}$ to form a $d^2 \times d^2$ unitary matrix.  Note the columns of $M$ are indexed from $0$ to $d^2-1$.  We will index the rows of $M$ using $ij$ pairs, consistent with the ordering of $B$; thus, row $0$ of $M$ corresponds to $i=0,j=0$; row 1, to $i=1,j=0$; and its final row, to $i=d-1, j=d-1$.  In general, row $jd+i$ has entries
\[   M_{ij,a} = \sqrt{\lambda_j} U_{ij}^{(a)} \text{ for } a=0,\ldots,K-1 \text{ and } M_{ij,b} = \phi_{ij}^{(b)} \text{ for } b=K,\ldots,d^2-1 . \]
Note that pairwise orthogonality among columns $a=0$ through $K-1$ of $M$ expresses orthogonality of the corresponding message states $\ket{\Psi^{(a)}}$ in $H$, or, equivalently, $\Lambda$ orthogonality of Alice's encoding matrices.  Of course, the rows of the unitary matrix $M$ also constitute an orthonormal set and hence necessarily satisfy
\begin{equation} \label{EG5}
   \sum_{a=0}^{K-1} M_{ij,a} (M_{i'j',a})^* + \sum_{b=K}^{d^2-1} M_{ij,b} (M_{i'j',b})^* \equiv \sqrt{\lambda_j \lambda_{j'}} \sum_{a=0}^{K-1} U_{ij}^{(a)} (U_{i'j'}^{(a)})^*  + \sum_{b=K}^{d^2-1} \phi_{ij}^{(b)} (\phi_{i'j'}^{(b)})^*   =  \delta_{i,i'} \delta_{j,j'}.
\end{equation}
Eq.~(\ref{EG5}) embodies inherent restrictions on the entries $U_{ij}^{(a)}$ of encoding unitaries, which lead to suprisingly simple derivations of previously proved dense coding results, as well as of hitherto unrecognized properties of $K$ as a function of the Schmidt coefficients.  It is the key relationship of the augmented message-matrix approach.  Our principal results from this approach are described and placed in context in the outline below.
\begin{itemize}
   \item Settling a conjecture by Mozes \textit{et al.}\ \cite{M},   Wu \textit{et al.}\  \cite{W} proved that for any dimension $d$ there is an upper bound 
 \begin{equation}\label{WCSGB}
   \lambda_0 \leq d/K 
  \end{equation}
  on the value of $\lambda_0$ permitting $K$ unitaries that are $\Lambda$ orthogonal, where $d \leq K \leq d^2$.  The derivation in \cite{W} of (\ref{WCSGB}), which we refer to as the WCSG bound on $\lambda_0$ (for $d$ and $K$),  rests on density-matrix manipulations. Subsequently Bourdon \textit{et al.} \cite{B} gave a derivation of (\ref{WCSGB}) which avoided the introduction of density matrices by utilizing  projection-operator techniques.   In Section~\ref{firstobs} below, we take an even more elementary approach, presenting an augmented message-matrix derivation of the WCSG bounds.
     \item Based primarily on numerical evidence, Mozes \textit{et al.}\ conjectured that there is no set of Schmidt coefficients that allow $K=d^2-1$; i.e., they conjectured that  whenever the state $\ket{\psi}$ of a two-qudit system supports transmission of $d^2-1$ messages via dense coding, then $\ket{\psi}$ is maximally entangled (and therefore $K = d^2$).     They proved this result analytically for $d=2$ only.  Ji \textit{et al.}\ \cite{J}  settled this conjecture for all $d$, utilizing partial trace techniques and the concavity of the von Neuman entropy of the entangled states.  Section~\ref{nod2m1} presents a simple proof for all $d$, utilizing straightforward manipulation of \EQ{EG5}.
     \item The numerical analysis of Mozes \textit{et al.}\ strongly suggests that when $K=d+1$,  $\lambda_0 \leq (d-1)/d$, rather than the less restrictive result $\lambda_0 \leq d/(d+1)$ that follows from (\ref{WCSGB}).   In Section~\ref{kedp1}, we show that when $K=d+1$, the bound $\lambda_0 \le d/K$  reduces to at least $\lambda_0 \leq 1/2(1+\sqrt{(d-2)/(d+2)})$, which is always less than $d/(d+1)$. 
   \item       Mozes \textit{et al.}\  showed that there is a state $\ket{\psi}$ with $\lambda_0 = (d-1)/d$ that supports $K=d+1$ messages by explicitly constructing $d+1$ encoding unitaries for $\ket{\psi}$, two of which are the identity $I$ and the shift $X$ (defined in section \ref{extensions} below.)  In Section~\ref{extensions},  we show that whenever one has a family of $d+1$ encoding unitaries, two of which are $I$ and $X$, then the WCSG bound $\lambda_0\le d/(d+1)$ is reduced all the way to $(d-1)/d$. (Note that one can always assume, without loss of generality, that a family of encoding unitaries includes $I$ (see, e.g., \cite[Lemma II.1]{B}).  Thus, the special assumption here is the inclusion of $X$ as well.)
      \item  We also provide in  Section~\ref{extensions} generalizations of the result just described, applicable when $K \ge d+1$ and the encoding unitaries include not only $I$ and $X$ but additional powers of $X$.  
       \end{itemize}

\section{\label{firstobs}First Observations} 

A particularly useful special case of Eq.~(\ref{EG5}) results from setting $i' = i$ and summing the result over $i$
\[  \sqrt{\lambda_j \lambda_{j'}} \sum_{a=0}^{K-1} \sum_{i=0}^{d-1} U_{ij}^{(a)} (U_{ij'}^{(a)})^*  + \sum_{b=K}^{d^2-1} \sum_{i=0}^{d-1}\phi_{ij}^{(b)} (\phi_{ij'}^{(b)})^*   =  d \delta_{jj'} , \]
which, after setting $j'=j$ and invoking the unitarity of the $U^{(a)}$'s, immediately yields
\begin{equation} \label{foeq}
  d - K \lambda_j = \sum_{b=K}^{d^2-1} \sum_{i=0}^{d-1}|\phi_{ij}^{(b)}|^2 \geq 0.
\end{equation}

From \EQ{foeq}, for the case $j=0$, comes rather trivially the WCSG bound result:
\[ \lambda_0 \leq \frac{d}{K} , \]
consistent with numerical results in Mozes \textit{et al.} \cite{M} and proven by other means in  \cite{W,B}.   \EQ{foeq} will later be utilized for other proofs in Section~\ref{nod2m1} and in Section~\ref{kedp1}.

Another observation, to be expanded upon later, involves the non-saturation of the WCSG bound when $K=d+1$, i.e. the impossibility of strict equality in the bound equation.   Bourdon {\it et al.}\ \cite[Proposition II.3]{B} established this non-saturation result, showing that whenever $\ket{\psi}$ of (\ref{EG1}) supports $K=d+1$ encoding unitaries, then $\lambda_0 < d/(d+1)$.   In order to illuminate the structure of augmented message-matrices, we present a short, simple proof of non-saturation under the assumption that all the initial state's Schmidt coefficients, namely $\sqrt{\lambda_j}$, $j=0, 1, \ldots, d-1$,  are nonzero.   

Suppose, in order to obtain a contradiction, that there exists a $\Lambda$-orthogonal family $\{U^{(a)}\}_{a=0}^{d}$ of $d+1$ encoding unitaries where $\lambda_0 = d/(d+1)$.   Let $M$ be a corresponding augumented message-matrix. We will assume $U^{(0)} = I$, without further loss of generality.  Because $\lambda_0 = d/(d+1)$, \EQ{foeq} shows entries $00$ through $(d-1)0$ in each of the $\phi^{(b)}$ vectors in $M$ to be zero.  Thus the remaining entries in each $\phi^{(b)}$ vector---those in rows $01$ through $(d-1)(d-1)$ of $M$---constitute a set of $d^2 - d -1$ orthonormal vectors in $\mathbb{C}^{d^2-d}$.   Let $W$ be the span of these vectors so that $W$ is a $d^2-d-1$ dimensional subspace of $\mathbb{C}^{d^2-d}$.  The ortho-complement $W^{\perp}$ of $W$ is thus one-dimensional; let $v$ be a unit vector spanning $W^{\perp}$.  For any $a\in \{0, 1, \ldots, d\}$,  let $v^{(a)}$ be the (necessarily nonzero) vector in $\mathbb{C}^{d^2-d}$ formed by the $01$ through $(d-1)(d-1)$ entries of column $a$ of $M$.  Orthogonality of columns of $M$ implies that $v^{(a)}$ belongs to $W^\perp$ for each $a$, so that  there are nonzero constants $\beta_a$ such that $v^{(a)} = \beta_a v$ for each $a\in \{0, 1, \ldots, d\}$.   Thus for each $a\in \{1,2, \ldots, d\}$, there is a nonzero constant $\gamma_a$ such that 
\begin{equation}\label{KE}
   v^{(a)} = \gamma_a v^{(0)}.
   \end{equation}
Observe that both $v^{(a)}$ and $v^{(0)}$ have length $1-\lambda_0$ so that $|\gamma_a|=1$.  Because $U^{(0)}=I$ and all Schmidt coefficients are nonzero, it follows from Eqs.~(\ref{EG3}) and (\ref{KE}) that $U_{ij}^{(a)}=\gamma_{a} \delta_{ij}$ for $0 \leq i \leq d-1$ and $1 < j \leq d-1$.  Therefore, because $|\gamma_{a}|=1$ and $U^{(a)}$ is unitary for each $a$, we further conclude that $|U_{00}^{(a)}|=1$.  Hence, employing Eq. (\ref{EG3}) with $a=0$ and, e.g. $b=1$, we obtain
$$
0 =   \text{tr}\left(\Lambda \left(U^{(0)}\right)^\dagger U^{(1)}\right) = \sum_{j=0}^{d-1} \lambda_j U^{(1)}_{jj} = \lambda_0 U^{(1)}_{00} + \gamma_1 \sum_{j=1}^{d-1} \lambda_j .
$$
Thus $\lambda_0 U^{(1)}_{00} = - \gamma_1(1-\lambda_0)$, which implies $|\gamma_1| = \lambda_0/(1-\lambda_0) = d > 1$, a contradiction.  From this we conclude that for $K=d+1$, when all Schmidt coefficents are nonzero, $\lambda_0$ must satisfy the strict inequality relationship
\[ \lambda_0 < \frac{d}{d+1} . \]

 In Section \ref{kedp1}, we drop the assumption that all Schmidt coefficients be nonzero and exhibit for every $d$ an upper limit for $\lambda_0$ that is a finite distance below the limit given by the WCSG bound $d/(d+1)$ for $K = d+1$.

\section{\label{nod2m1}Impossibility of only $d^2-1$ encoding unitaries in $d$ dimensions} 

It is known that under the $d$-dimensional deterministic dense coding protocol there is a region of the space of the $\lambda_j$ that admits a maximum of $K=d$ encoding unitaries (coinciding with the limit of classical communication) as well as a second region (actually no more than a point) wherein one can find as many as $K=d^2$ encoding unitaries, the maximum number possible in view of the fact that our initial $\ket{\psi}$ of Eq.~(\ref{EG1}) lies in a $d^2$-dimensional Hilbert space.  One expects, therefore, that there should be regions of the $\lambda_j$ space wherein $K=m$ but no more than $m$ encoding unitaries can be found, for every integer value of $m$ from $m=d+1$ to $m=d^2-1$.  One of the more counterintuitive properties of the  protocol is that this just-stated expectation is met for every such $m$ {\textbf except} $m = d^2-1$.  The only point in the space of $\lambda_j$'s that allows $d^2 - 1$ encoding unitaries also allows $d^2$, occuring at the point where $\lambda_j = 1/d$ for all $j$.  This result was proven by Ji \textit{et al.} \cite{J}, using the spectral properties of partial traces of density operators and the concavity of the von Neuman entropy.  Our approach allows this result to be established algebraically from the unitarity properties of the encoding $U^{(a)}$'s and the corresponding augmented message-matrix $M$ introduced in Section~\ref{iandf}.

Consider the $d$-dimensional situation, with $K=d^2-1$ encoding unitaries $\{U^{(a)}\}_{a=0}^{d^2-2}$.  From the WCSG bound equation of Section~\ref{firstobs} we know in this case  $\lambda_0 \leq d/(d^2-1)$, and it is easily seen that this ensures that all the $\lambda_i$ are non-zero. The sum over $a$ in \EQ{EG5} involves $d^2-1$ terms, and the sum over $b$ has only a single term allowing us to drop the superscript $b = d^2-1$ on $\phi$ for this case.  
Extract from \EQ{EG5} the following relationship by setting  $j'=j$, dividing through by $\lambda_{j}$, and summing the result over $j$:
\begin{equation} \nonumber
   \delta_{ii'} \sum_{j=0}^{d-1} \frac{1}{\lambda_{j}}  =  \sum_{a=0}^{d^2-2} \sum_{j=0}^{d-1} U^{(a)}_{ij} (U^{(a)}_{i'j})^* + \sum_{j=0}^{d-1} \frac{\phi_{ij} \phi_{i'j}^*}{\lambda_{j}}   .\end{equation}
Since each of the $d^2-1$ $U^{(a)}$'s are unitary, we reach
\begin{equation}\label{p7l}
   \sum_{j=0}^{d-1} \frac{\phi_{ij} \phi_{i'j}^*}{\lambda_{j}} = \left( \sum_{j=0}^{d-1} \frac{1}{\lambda_{j}} - (d^2-1) \right) \delta_{ii'} .
\end{equation}
This equation can be interpreted as describing the properties of a  $d \times d$ matrix $S$, with its $i,j$ element equal to $\phi_{ij}/ \sqrt{\lambda_j}$, whose rows (labeled by $0 \leq i \leq d-1$) are vectors which are mutually orthogonal, and each of which has a common length, the square of which may be computed by setting $i'=i$ in (\ref{p7l}):
\begin{equation} \nonumber
   \sum_{j=0}^{d-1} \frac{|\phi_{ij}|^2}{\lambda_j} = \sum_{k=0}^{d-1} \frac{1}{\lambda_k}  - (d^2-1)   .
\end{equation}
Note that the value of the common row-lengths must be nonzero; otherwise all components $\phi_{ij}$ of the unit vector $\phi$ would be zero, a contradiction.  Such a matrix, within a constant multiple of a unitary matrix, will have its columns (labeled by $0 \leq j \leq d-1$) representable as vectors with the same lengths as the row vectors; thus
\begin{equation} \label{eqc}
   \sum_{i=0}^{d-1} \frac{|\phi_{ij}|^2}{\lambda_j} = \sum_{k=0}^{d-1} \frac{1}{\lambda_k} - (d^2-1)   .
\end{equation}
From \EQ{foeq} we arrive at another expression for the squares of these lengths (recalling we have dropped the superscript $b=d^2-1$ on $\phi$):
\begin{equation} \label{eqd}
   \sum_{i=0}^{d-1} \frac{|\phi_{ij}|^2}{\lambda_j} = \frac{d}{\lambda_j} - (d^2-1)  . 
\end{equation}
In combination, Eq.'s (\ref{eqc}) and (\ref{eqd}) produce
\[ \sum_{k=0}^{d-1} \frac{1}{\lambda_{k}} = \frac{d}{\lambda_j} , \]
true for each $j$, $0 \leq j \leq d-1$.  This is only possible if all the $\lambda_j$'s are equal, and since they sum to one, necessarily $\lambda_j = 1/d$ for each $j$.  This corresponds to the set of Schmidt coefficients for maximal entanglement, the point at which $d^2$ encoding unitaries can be found.  Thus we have proved that if we can find $d^2-1$ encoding unitaries, we must be able to find $d^2$, i.e., there is no region in the space of the $\lambda_j$ that admits a maximum of $K=d^2-1$ encoding unitaries.

Evaluating the expression for the common length of the row and column vectors of the matrix $S$,
\[ \sum_{k=0}^{d-1} \frac{1}{\lambda_{k}}  - (d^2-1) = \sum_{k=0}^{d-1} d  - (d^2-1) = 1 , \]
and because (as previously noted) the rows of $S$ are orthogonal, this shows $S$ to be unitary, and, in fact, to be the last encoding unitary matrix:
\[ U^{(d^2-1)}_{ij} = S_{ij} = \frac{\phi_{ij}}{ \sqrt{\lambda_j}}. \]

\section{\label{kedp1}Bound for the $K=d+1$ case} 

 The WCSG bound for the case $K = d+1$ is  $\lambda_0 \le d/(d+1)$.   A smaller bound, $\lambda_0 \le (d-1)/d$, has been conjectured by Mozes \textit{et al.} \cite{M} based on the failure of a numerical-search procedure to find families of $d+1$ encoding unitaries when the value of  $\lambda_0$ exceeds $(d-1)/d$.  Also providing some support of their conjecture,  Mozes \textit{et al.} have, for all $d$, analytically constructed families of $d+1$ encoding unitaries for $\lambda_0 = (d-1)/d$ and $\lambda_j = 0$ for $j\ge 2$.  Using the augmented message-matrix, we establish here an upper bound on $\lambda_0$ for the case $K=d+1$ that is a finite distance below the WCSG bound of $d/(d+1)$ but not as small as the conjectured bound $(d-1)/d$.    Throughout this section, we assume that $\lambda_0$ is confined to the region of interest:  $(d-1)/d \le \lambda_0 \le d/(d+1)$.  We focus our attention on dimensions $d$ higher than two, the one case in which the conjectured bound has been proven.  Thus, in particular, if $\lambda_0$ lies in our region of interest, then $\lambda_0 > 1/2$.

Assume that Alice can create a maximum of $K = d+1$ distinguishable messages. We assume $U^{(0)}$ to be the identity operator, so that the entries in the first column of the augmented message-matrix $M$ are
\[ \me{ij}{M}{0} =  \sqrt{\lambda_j} U^{(0)}_{ij} = \sqrt{\lambda_j} \delta_{ij} . \]
$\Lambda$ orthogonality of each $U^{(a)}$, $1 \leq a \leq d$, with $U^{(0)}$, which is equivalent to the the orthogonality of columns $a$ and $0$ of $M$, leads to
\begin{equation}
    \sum _{i=0}^{d-1}  \sum _{j=0}^{d-1} \me{ij}{M}{a} \me{ij}{M}{0}^{*} = \sum _{i=0}^{d-1} \lambda_i U^{(a)}_{ii} = 0 .
\end{equation}
Define real constants $\eta_i \equiv \lambda_i/\lambda_0$ for $1 \leq i \leq d-1$, and use them to write
\[  |U^{(a)}_{00}| = \left| \sum_{i=1}^{d-1} \eta_i U^{(a)}_{ii} \right| \leq \sum_{i=1}^{d-1} \eta_i |U^{(a)}_{ii}|, \;\; 1\leq a \leq d,  \]
where the final step follows from the triangle inequality.  From this it follows simply that
\[  \sum _{a=0}^{d}  |U^{(a)}_{00}|^2 = 1+\sum_{a=1}^d |U^{(a)}_{00}|^2 \leq 1+\sum_{a=1}^d \left( \sum_{i=1}^{d-1} \eta_i |U^{(a)}_{ii}| \right)^2 . \]
Reordering the summations in the rightmost term of this expression we see
\[  \sum_{a=1}^{d} \left( \sum_{i=1}^{d-1} \eta_i |U^{(a)}_{ii}| \right)^2 =
     \sum _{i=1}^{d-1}  \sum _{j=1}^{d-1}  \eta_i \eta_j \left( \sum_{a=1}^{d} |U^{(a)}_{ii}| |U^{(a)}_{jj}| \right) . \]
Let $q \geq 1$ be the value of $i$ that maximizes $\sum_{a=1}^{d} |U^{(a)}_{ii}|^2$ for $1 \leq i \leq d-1$; then
\[  \sum_{a=1}^d | U^{(a)}_{ii}| \; |U^{(a)}_{jj}| \leq \frac12 \sum_{a=1}^d(|U_{ii}^{(a)}|^2+|U_{jj}^{(a)}|^2) \leq \sum_{a=1}^d |U^{(a)}_{qq}|^2, \]
which leads to  
\begin{equation} \sum_{a=0}^d | U^{(a)}_{00}|^2 \leq 1+ R_q  \sum _{i=1}^{d-1}  \sum _{j=1}^{d-1}  \eta_i \eta_j = 1 + \eta^2 R_q \label{uoos} , \end{equation}
where we have defined 
\[ \eta \equiv \sum_{i=1}^{d-1} \eta_i = \frac{1-\lambda_0}{\lambda_0} \text{     and     } R_q \equiv \sum_{a=1}^d |U^{(a)}_{qq}|^2 . \]
Using $U^{(0)} = I$ and the fact that length of each row of  $U^{(a)}$ is one, we have
\[ \sum_{a=0}^d | U^{(a)}_{q0}|^2 = \sum_{a=1}^d | U^{(a)}_{q0}|^2 = \sum_{a=1}^d \left( 1 - \sum_{k=1}^{d-1}| U^{(a)}_{qk}|^2  \right) \leq d - R_q .\]
Normalization of the $(q0)^{th}$ row of $M$ allows the previous equation to be transformed into
\begin{equation} \label{qrow}
  1 - \sum_{b=d+1}^{d^2-1} |\phi_{q0}^{(b)}|^2  = \lambda_0 \sum_{a=0}^d | U^{(a)}_{q0}|^2 \leq \lambda_0 (d-R_q) .
\end{equation}
From Eq.~(\ref{uoos}),
\[ R_q \geq \frac{ \sum_{a=0}^d |U^{(a)}_{00} |^2 - 1}{\eta^2} , \]
which along with normalization of the $(00)^{th}$ row of $M$,
\[ \lambda_0 \sum_{a=0}^d |U^{(a)}_{00}|^2 + \sum_{b=d+1}^{d^2-1} |\phi_{00}^{(b)}|^2 = 1 ,\]
yields
\begin{equation}
  R_q \geq \frac{ 1 - \lambda_0 - \sum_{b=d+1}^{d^2-1} |\phi_{00}^{(b)}|^2}{\lambda_0 \eta^2}.
 \end{equation}
 Combining this with Eq.~(\ref{qrow}) gives
\[ 1 - \sum_{b=d+1}^{d^2-1} |\phi_{q0}^{(b)}|^2  \leq \lambda_0 \left( d-\frac{ 1 - \lambda_0 - \sum_{b=d+1}^{d^2-1} |\phi_{00}^{(b)}|^2}{\lambda_0 \eta^2} \right), \]
and thus 
\begin{equation} \label{oneminusG}
 1 - d \lambda_0 + \frac{\lambda_0^2}{1-\lambda_0} \leq \sum_{b=d+1}^{d^2-1}|\phi_{q0}^{(b)}|^2  + \frac{\lambda_0^2}{(1-\lambda_0)^2} \sum_{b=d+1}^{d^2-1}|\phi_{00}^{(b)}|^2 . 
\end{equation}
Applying (\ref{foeq}) with $j=0$ and $K=d+1$, we obtain
\begin{equation}
  \label{AGEsum2} \sum_{b=d+1}^{d^2-1} \sum_{k=0}^{d-1} |\phi^{(b)}_{k0}|^2  = d - (d+1) \lambda_0 .
\end{equation}
Eq.~(\ref{AGEsum2}) along with Eq.~(\ref{oneminusG}) yields the following sequence of inequalities
\begin{eqnarray}
 1 - d \lambda_0 + \frac{\lambda_0^2}{1-\lambda_0} &\leq& \sum_{b=d+1}^{d^2-1} |\phi^{(b)}_{q0}|^2 + \frac{\lambda_0^2}{(1-\lambda_0)^2} \left( d - (d+1) \lambda_0 - \sum_{b=d+1}^{d^2-1} \sum_{k=1}^{d-1} |\phi^{(b)}_{k0}|^2 \right)  \nonumber \\ \label{AGEsum4}
 &\leq& \sum_{b=d+1}^{d^2-1}|\phi_{q0}^{(b)}|^2  \left(1 - \frac{\lambda_0^2}{(1-\lambda_0)^2} \right) + (d - (d+1) \lambda_0) \frac{\lambda_0^2}{(1-\lambda_0)^2},
\end{eqnarray}
where to obtain (\ref{AGEsum4}), we have used $\sum_{k=1}^{d-1}|\phi_{k0}^{(b)}|^2  \geq |\phi_{q0}^{(b)}|^2$, which is obvious since $q \geq 1$.
Note that in the region of interest, $\lambda_0 > 1/2$, the term $1 - \frac{\lambda_0^2}{(1-\lambda_0)^2}$ will be negative; thus we maximize the RHS of the last inquality by inserting the smallest possible value of $\sum_b |\phi_{q0}^{(b)}|^2 $, which is zero, leading to 
\[ 1 - d \lambda_0 + \frac{\lambda_0^2}{1-\lambda_0} \leq (d - (d+1) \lambda_0) \frac{\lambda_0^2}{(1-\lambda_0)^2 }. \]
Solving for $\lambda_0$ produces
\begin{equation} \lambda_0 \leq \frac12 \left( 1+\sqrt{\frac{d-2}{d+2}} \right)  , \label{lower} \end{equation}
which is strictly less than the WCSG bound of $d/(d+1)$ for $d>2$.
For $d=3$ it gives $\approx 0.7236$, less than the WCSG bound of $3/4$, but only about one-third of the way towards the $2/3$ bound conjectured by Mozes {\textit et al}.  

It might be thought that even when $K$ is not restricted to the value $K=d+1$, an argument similar to the one just given also would push down, towards smaller values of $\lambda_0$, the WCSG bound $\lambda_0 \leq d/K$.  Saturation has been demonstrated in \cite{B} for the WCSG bound for $K=d+2$ and $K=2d-1$.  Generalization of the argument of this section for other $K$ values is easily constructed, but shows that an improvement in the WCSG bound comes only for the case $K=d+1$ shown above.

\section{\label{extensions}Extension results in case $K = d+1$ } 

In the preceding section, we established that when $K = d+1$, the WCSG bound $\lambda_0\le d/(d+1)$ can be reduced to the value given in Eq.~(\ref{lower}).    However, as we have indicated, Mozes \textit{et al.} \cite{M} have conjectured that this bound can be further reduced to $(d-1)/d$; moreover, they have constructed for every $d$ a family of $d+1$ encoding unitaries when $\lambda_0 = (d-1)/d$.   These families of $d+1$ encoding unitaries include both the identity operator $I$ and the shift operator $X$ defined by
\[ X \ket{j} \equiv \ket{j\!+\!1}, \;\;\;\; j=0,1,2, \ldots,d\!-\!1, \text{   where } \ket{d} \equiv \ket{0}   . \]
In this section, we show that whenever there is a family of $d+1$ encoding unitaries that includes $I$ and $X$, then  $\lambda_0 \le (d-1)/d$, in agreement with Mozes' conjecture.  We also obtain a more general bound on $\lambda_0$ in cases where an encoding family of unitaries includes not only $I$ and $X$ but additional powers of $X$.

We again address the $d$-dimensional dense coding problem, with $K = d+1$.  We seek the conditions under which we can have a set of $d+1$ encoding unitaries $\{U^{(a)}\}_{a=0}^{d}$ that include both the identity, $I \equiv U^{(0)}$, and $X \equiv U^{(1)}$.  Note that any set of encoding unitaries can be transformed so as to include $I$; assuming that $X$ is also included constitutes a special case.   

For the remaining $d-1$ unitaries, $\{U^{(a)}\}_{a=2}^{d}$, their $\Lambda$ orthogonality with the identity requires 
\begin{equation} U^{(a)}_{00} = - \sum_{j=1}^{d-1} \eta_j U^{(a)}_{jj} \label{ieta} \end{equation}
and their $\Lambda$ orthogonality with $X$ requires
\begin{equation} U^{(a)}_{10}  = - \sum_{j=1}^{d-1} \eta_j U^{(a)}_{(j\!+\!1)j}  , \label{xeta} \end{equation}
recalling the definition $\eta_j \equiv \lambda_j / \lambda_0.$  (Indices for matrix elements of $U^{(a)}$'s should be interpreted as integers modulo $d$, thus a term like  $ \me{d}{U^{(a)}}{d-1}$, should be identified with $\me{0}{U^{(a)}}{d\!-\!1}$.)  
Unitarity of each $U^{(a)}$ ($a>1$) allows us to write
\[ \sum_{j=2}^{d-1} |U^{(a)}_{j0} |^2 = 1 - \left|U^{(a)}_{00} \right|^2 - \left|U^{(a)}_{10} \right|^2 = 
1 - \left|\sum_{j=1}^{d-1} \eta_j U^{(a)}_{jj}\right|^2 - \left| \sum_{j=1}^{d\!-\!1} \eta_j U^{(a)}_{(j\!+\!1)j}\right|^2 . \]
For $a\in \{2, 3, \ldots, d\}$  and $j\in \{2, 3, \ldots, d-1\}$,  let $v^{(a)}_j$ be the two-dimensional vector whose first component is $U^{(a)}_{jj}$ and second is $U^{(a)}_{(j\!+\!1)j}$, with length  $\|v^{(a)}_j\| \le 1$ since this vector consists of two components of one of the columns of  the unitary matrix $U^{(a)}$. We then can re-express the previous equation as
\begin{eqnarray} 
\sum_{j=2}^{d-1} |U^{(a)}_{j0} |^2 & = & 1 - \left\| \sum_{j=1}^{d-1} \eta_j v^{(a)}_j \right\|^2 \nonumber\\      
                     &\ge & 1 - \left(\sum_{j=1}^{d-1} \eta_j\|v^{(a)}_j\|   \right)^2 \nonumber \\                                             
                     & \ge &1- \left( \sum_{j=1}^{d-1} \eta_j \right)^2 \nonumber \\
                     & = & 1-\eta^2,\label{TRI}
\end{eqnarray} 
where we have used the triangle inequality to establish the second line of (\ref{TRI}), and used $\eta \equiv \sum_{j=1}^{d} \eta_j = (1-\lambda_0)/\lambda_0$ in the final step.

Starting with Eq.~(\ref{EG5}), setting $j = j' = 0$, $i' = i$, and summing over $i$ from $2$ to $d-1$ produces
\[ d-2 = \lambda_0 \sum_{a=0}^{d} \sum_{i=2}^{d-1} |U^{(a)}_{i0} |^2 + \sum_{b=d+1}^{d^2-1}\sum_{i=2}^{d-1} |\phi_{i0}^{(b)}|^2. \]
The $a=0$ and $a=1$ terms in the first sum are both zero, therefore (using \EQ{TRI})
\[ d-2 \geq \lambda_0 \sum_{a=2}^{d} \sum_{i=2}^{d-1} |U^{(a)}_{i0} |^2 \geq \lambda_0 (d-1)(1-\eta^2) \]
and consequently, $\lambda_0 \leq (d\!-\!1)/d.$  Thus, the inclusion of both $I$ and $X$ as encoding unitaries in the $K=d+1$ case leads to the upper bound on $\lambda_0$ as postulated by Mozes \textit{et al.} for the more general case. 

\textit{Remarks.}  
\begin{list}{}
   \item{(a)} The argument presented above is easily modified to show that if a family of $d+1$ encoding unataries contains both $I$ and $X^j$ for some $j  \in \{1,2,\ldots,d-1\}$ then $\lambda_0 \leq (d-1)/d$.  Note that  if an encoding family $\{U^{(a)}\}$ contains any two operators from $\{X^k\}_{k=0}^{d-1}$, say $U^{(0)} = X^n$ and $U^{(1)} =X^m$, where $n >m$, then  $\{X^{d-n}U^{(a)}\}$ will be an encoding family as well, one that includes $I$ and $X^j$, where $j = d-n+m$.  Thus, $\lambda_0 \le (d-1)/d$ whenever there is an encoding family of $d+1$ unitaries containing any two (distinct) elements from $\{X^k\}_{k=0}^{d-1}$.
   \item{(b)} A more careful analysis of the inequalities we have used in Eq.~\ref{TRI} to derive the bound $\lambda_0 \le (d-1)/d$ leads to the following improvement: If there exists a $\Lambda$-orthogonal family of d+1 encoding unitaries including both $I$ and $X$ and if $\lambda_2 >0$, then $\lambda_0 < (d-1)/d$, an inequality also consistent with numerical results obtained by Mozes \textit{et al.} 
\end{list}
The method described in this section is quite general and can quickly be adapted to produce a more general result.  Assume that $I$ and $m-1$ additional powers of $X$ are the first $m$ encoding unitaries in the family $\{U^{(a)}\}_{a=0}^{K-1}$, with $m \ge 1$, $d \ge m$, and $d+1 \leq K \leq d^2$.  There will be $m$ expressions of the form of (\ref{ieta}) and (\ref{xeta}) that simply lead to a generalization of (\ref{TRI}):
\[ \sum_{j=m}^{d-1} |U^{(a)}_{j0} |^2  \geq 1-\eta^2. \]
The key equation of the augmented message-matrix approach, Eq.~(\ref{EG5}), with $j = j' = 0$, $i' = i$, and summed over $i$ from $m$ to $d-1$ produces
\begin{equation} d-m \geq \lambda_0 (K-m) (1-\eta^2) \nonumber \end{equation}
leading to 
\begin{equation}   \lambda_0 \le \frac{K -m}{2K - m - d}. \label{genrl} \end{equation}
 Since $m \leq d$, the limit given on the right side of Eq. (\ref{genrl}) is no less than $1/2$.  It can therefore not produce a useful, stricter limit than the WCSG bound of $d/K$ unless $d/K \geq 1/2$, thus $K \leq 2d$.

Examples utilizing this result include:
\begin{list}{}
   \item{(a)} if $m = 2$ and $K = d+1$, we reproduce the result proved above, i.e.  $\lambda_0 \le (d-1)/d$;
   \item{(b)} if $m = 3$, $d=4$, and $K = 5$, then $\lambda_0 \le 2/3$; and
   \item{(c)} if $m = 2$ and $K = d+2$, we get $\lambda_0 \le d/(d+2)$.
\end{list}
It is also noteworthy that if $m=d$, then $\lambda_0 \leq 1/2$ proving by an entirely different method Proposition III.2 from our previous work \cite{B}.

We conclude by showing the bound (\ref{genrl}) is not always saturated by exhibiting a more restrictive bound for the case $m=2$, and $K = d+2$, under the additional requirements that $d = 3$ and $\lambda_2 = 0$.  Thus we seek conditions under which there is a family of five $3 \times 3$ encoding unitaries $\{U^{(a)}\}_{a=0}^{4}$ where $U^{(0)}=I$, $U^{(1)}=X$, and $\lambda_2=0$.
$\Lambda$ orthogonality among the three unspecified unitaries, $U^{(a)}, \; a=2,3,4,$ involves only the first two columns of each, so for each $U^{(a)}$  we will focus only on the six elements from those columns.  $\Lambda$ orthogonality of each $U^{(a)}$ with $I$ requires $U^{(a)}_{00} = - \eta U^{(a)}_{11}$ and $\Lambda$ orthogonality with $X$ requires $U^{(a)}_{10} = - \eta U^{(a)}_{21}$.  Focusing on the six non-negative quantities $T_{ij} \equiv \sum_{a=2}^4 |U^{(a)}_{ij}|^2; \; i=0,1,2; \; j=0,1$, the prior requirements imply $T_{00} = \eta^2 T_{11}$ and $T_{10} = \eta^2 T_{21}$.  Furthermore, the requirement that each column of $U^{(a)}$ is a unit vector implies that $T_{20} = 3 - T_{00} - T_{10} = 3 - \eta^2(T_{11}+T_{21})$; and $T_{01} = 3 - T_{11}-T_{21}$.  We now proceed to establish a series of inequalities that restrict values of the two free $T_{ij}$ values which we pick to be $T_{11}$ and $T_{21}$.

To establish relations based upon the unitarity of the augmented message-matrix, consider Eq.~(\ref{EG5}).  Setting $i'=i$ and $j'=j$ we get the following set of inequalities based upon the requirements of unit length of the rows:
\begin{eqnarray}
   \lambda_0 (1 + T_{00}) = \lambda_0 ( 1 + \eta^2 T_{11}) \leq 1 \label{equ1} \\
   \lambda_0 (1 + T_{10}) = \lambda_0 ( 1 + \eta^2 T_{21}) \leq 1 \label{equ2} \\
   \lambda_0 ( T_{20}) = \lambda_0 ( 3 - T_{00} - T_{10})=\lambda_0 ( 3 - \eta^2(T_{11} + T_{21}))\leq 1 \label{equ3} \\
   \lambda_1 (T_{01}) = \lambda_0 ( 3 - T_{11} - T_{21}) \leq 1 \label{equ4} \\
   \lambda_1 (1 + T_{11}) \leq 1 \label{equ5} \\
   \lambda_1 (1 + T_{21}) \leq 1 \label{equ6} .
\end{eqnarray}
It is easy to demonstrate that the last two of these inequalities are redundant with the first two.  Other restrictions come from the requirements that the rows of the $U^{(a)}$'s are unit vectors:
\begin{eqnarray}
   T_{00} + T_{01} = \eta^2 T_{11} + 3 - T_{11} - T_{21} \leq 3 \label{equ7} \\
   T_{10} + T_{11} = \eta^2 T_{21} + T_{11} \leq 3 \label{equ8} \\
   T_{20} + T_{21} =  3 - \eta^2 T_{21} - \eta^2 T_{11}  + T_{21} \leq 3 \label{equ9} .
\end{eqnarray}
The first of these is satisfied trivially in our region of interest, $\eta^2 \leq 1$.  There remain six relations that must be simultaneously satisfied.  Of these, three control the possibility of any acceptable solution in the region of interest, $0.6 \leq \eta \leq 1$:  Equations (\ref{equ1}, \ref{equ3}, and \ref{equ9}).  Observe that (\ref{equ1}) is equivalent to $T_{11} \leq 1/\eta$; subsituting this into (\ref{equ9}) yields $T_{21} \leq \eta / (1-\eta^2)$; and subsituting these two inequalities into (\ref{equ3}) produces
\[ 2(1-\eta)- \frac{\eta^3}{1-\eta^2} \leq 0, \]
yielding $\eta \approx .68889$.  Thus, while $K = 5$ solutions can be found generally for values of $\lambda_0 \leq 3/5$ ($\eta \geq 2/3$) \cite[Proposition III.5]{B}, we have shown that if both $I$ and $X$ are included as encoding unitaries, a more restrictive limit applies --  $\lambda_0 \leq \approx.5921$ ($\eta \geq \approx.68889$).  Following this process there are likely to be a variety of other restrictive limits that can be established.


\end{document}